\documentclass{article}
\usepackage[utf8]{inputenc}
\usepackage{graphics}
\usepackage{graphicx}
\usepackage[margin=2cm]{geometry}
\usepackage{caption}
\usepackage{subcaption}
\usepackage{color}
\usepackage{amsmath}
\usepackage{amssymb}

\newcommand{\nd}{\noindent}

\title{Multicomponent holographic dark energy model with generalized entropy}
\author{\small{Facundo Abaca$^{1}$, Dario Javier Zamora$^{2,3}$\thanks{E-mail: dariojavier.zamora@uninsubria.it}}, \\
\small{$^1$ Departamento de F\'isica, Facultad de Ciencias Exactas y Tecnolog\'ia, Universidad Nacional de Tucum\'an}\\
\small{ Av. Independencia 1800, Tucuman, CP 4000, Argentina}\\
\small{$^2$ Dipartimento di Scienza e Alta Tecnologia, Universit{\`a} degli Studi dell’Insubria, Via Valleggio 11, 22100 Como, Italy}\\
\small{$^3$ Instituto de Fisica del Noroeste Argentino, CONICET and Universidad Nacional de Tucum\'an}
}

\date{\today}

\begin{document}

\maketitle

\begin{abstract}
%1 sentence each:
%What is the topic/definitions
%What is the problem/motivation
%What we made
%What is the novelty
%What we obtained/results
%Future perspectives/utility/applications

Holographic dark energy cosmology, also known as entropic cosmology, provides a concrete physical understanding of the late accelerated expansion of the universe. The acceleration appears to be a consequence of entropy associated with information storage in the universe. Therefore, the assumption of an ad-hoc dark energy is not necessary. In this paper, we investigate the implications of a multicomponent model (radiation and non-relativistic matter) that includes a subdominant power-law term within a thermodynamically admissible model. We use a generic power-law entropy and the temperature of the universe horizon results from the requirement that the Legendre structure of thermodynamics is preserved. We analyse the behaviour for different combinations of the parameters and compare them with other cosmological models, the observed redshift dependencies of the Hubble parameter $H$ and the luminosity distance data obtained from supernovae. This is an early attempt to analyse a multicomponent holographic dark energy model. Furthermore, the analysis is based on a entropy scaling with an arbitrary power of the Hubble radius instead of a specific entropy. This allows us to simultaneously infer different models, compare them and conserve the scaling exponent as a parameter that can be fitted with the observational data, thus providing information about the form of the actual cosmological entropy and temperature. We show that the introduced correction term is able to explain different acceleration and deceleration periods in the late-time universe by solving the model numerically. We discuss the advantages and disadvantages of holographic dark energy models compared to mainstream cosmology.

\end{abstract}

\section{Introduction}

%DEFINITION OF CONCEPTS AND CONTEXT OF THE TOPIC
%STATE OF THE ART
%HOW THE PAPER IS STRUCTURATED

In mainstream cosmology, matter and spacetime emerged from a singularity and evolved in four distinct periods, namely early inflation, radiation, dark matter and late-time expansion. During the phases dominated by radiation and dark matter, the universe is decelerating while the early and late-time expansion are accelerating stages.
The Lambda Cold Dark Matter ($\Lambda$CDM) model assumes a cosmological constant $\Lambda$ and the existence of a mysterious component called dark energy. This model is the simplest that can explain an accelerated expansion of the late universe, which according to the $\Lambda$CDM model is driven by dark energy. However, it implies some theoretical peculiarities, such as the cosmic coincidence and the problem of the cosmological constant \cite{Weinberg1989,Carroll2001}.
A possible connection between the acceleration periods is still unknown, and interestingly, the most popular candidate for dark energy driving the current acceleration phase ($\Lambda$- vacuum) relies on the cosmological constant and coincidence puzzle. To cope with these difficulties, several alternative models have been proposed, see for example \cite{Weinberg2008,Ellis2012,Komatsu2014,Sola2013}.

In this context, the holographic dark energy (HDE), also called entropic cosmology, is an interesting attempt that can address this problem using the holographic hypothesis. Horizon entropy is the backbone of HDE models, and therefore any change in the horizon entropy affects the HDE model. Various scenarios can be found, some of which are consistent with various astronomical observations \cite{Zhang2005, Zhang2007, Huang2004, Enqvist2005, Shen2005}. From this point of view, the controversial dark energy component is not necessary. Here, the accelerated expansion in the late period is based on the concept of entropic- force. Instead of dark energy, we have the holographic principle and entropy as the source of the late acceleration phase of the universe. An entropic force is an emergent phenomenon that results from the natural tendency of a thermodynamic system to increase its entropy, rather than from any particular underlying fundamental force. There is no field associated with an entropic force. The force equation is expressed in terms of a spatial dependence of the entropy $S$. The cosmological entropic force $F$, is then given by

\begin{equation}
F=-T\frac{dS}{dr_H},
\label{F}
\end{equation}

\nd where $r_H$ is the Hubble radius.

Thermodynamic properties of the universe have always attracted attention \cite{Davies1987,Prigogine1988,Prigogine1989} and in recent years especially entropic cosmology, see \cite{Cai2010,Karami2011,Mitra2015,Viaggiu2015} and citations therein.
Here we follow the attempts of Easson, Frampton and Smoot (EFS) \cite{Easson2011,Easson2012}. The first EFS model \cite{Easson2011} assumes that the entropy and temperature associated with the horizon of the universe are the Bekenstein-Hawking entropy \cite{Bekenstein1973} and the Hawking temperature \cite{Hawking1974}, respectively.
Since gravity is a long-range interaction, one can also use generalised statistical mechanics to study gravitational systems \cite{Komatsu2014,Majhi2017,Abe2001,Touchette2002,Sayahian2018,Moradpour2018,Komatsu2017,Moradpour2017,Moradpour2016,Abreu2013,Abreu2013a,Barboza2015,Nunes2016,Komatsu2013,Komatsu2014a,Komatsu2016,Tsallis2013}. This entropy was proposed in \cite{Tsallis2013} in the context of black holes. We would like to remind the reader that for such systems the additive Bekenstein-Hawking entropy is proportional to the area of the horizon, i.e. it is subextensive, while the non-additive $S_{\delta=3/2}$- entropy is proportional to the volume (at least in the case of equal probabilities), i.e. it is extensive, as required by thermodynamics. In the models \cite{Easson2011, Easson2012, Komatsu2013}, the expression for the temperature of the Hubble horizon is not derived from a clean physical principle. It is simply assumed to be the Hawking temperature expressed by the parameters of the universe, namely

\begin{equation}
T_{BH}(t)=\frac{\hbar c}{2\pi k_Br_H(t)}=\frac{\hbar H(t)}{2\pi k_B},
\label{Thawking}
\end{equation}

\nd where $c$ is the speed of light, $\hbar$ is the reduced Planck constant, $k_B$ the Boltzmann constant, and $H(t)$ is the Hubble parameter. $H$ is defined as 

\begin{equation}
H\equiv\frac{c}{r_H}=\frac{\dot{a}}{a},
\end{equation}

\nd $a=a(t)$ being the scale factor.

Arbitrary combinations of entropy and temperature could violate the Legendre structure of thermodynamics. This is not the case with the Bekenstein-Hawking entropy and the Hawking temperature, as proposed by EFS \cite{Easson2011}. This problem was recently discussed in \cite{Zamora2022}, where a physical principle for deriving the thermodynamically consistent temperature associated with each class of entropy was proposed. More precisely, the temperature must be derived from the Legendre structure of thermodynamics

\begin{equation}
\begin{split}
G(V,T, p,\mu,...)&=U(V,T,p,\mu,...)-TS(V,T,p,\mu,...)\\
&+pV-\mu N(V,T,p,\mu,...)-...,
\end{split}
\label{Legendre}
\end{equation}

\nd where $T,p,\mu$ are the temperature, pressure, and chemical potential, and $U,S,V,N$ are the internal energy, entropy, volume, and the number of particles of the system, respectively. This implies that, as detailed in \cite{Tsallis2013}, in a Schwarzschild (3+1)-dimensional black hole, the relation

\begin{equation}
\theta=1-d
\label{key2}
\end{equation}

\nd must hold, where $d$ is the dimension ($S\propto L^d$), and $\theta$ is the corresponding exponent for the scaling of the temperature ($T\propto L^\theta$), $L$ being a characteristic linear dimension of the d-dimensional system. For the Bekenstein-Hawking entropy ($d=2$), the temperature depends on $L^{-1}$ as the Hawking temperature.

The entropic force term, Eq. (\ref{F}), affects the background evolution of the late universe; in the present paper we do not focus on the inflation of the early universe. It has been shown that models with entropic forces containing only $H^2$ terms are not able to describe both decelerating and accelerating phases on a single basis \cite{ Zamora2022, Perico2013,Komatsu2013b, Basilakos2012}. Nevertheless, changes in the deceleration parameter can be smoothly introduced by including subdominant terms in the entropy of the horizon \cite{Easson2012,Zamora2022b}. Subdominant entropic terms are not rare in nature, even if they are neglected in first approximations. Moreover, they have a physical meaning and their inclusion is physically motivated. With this additional term, the model predicts different periods of acceleration and deceleration. With this in mind, we investigate here the effects of adding a subdominant power-law term in the entropy and simultaneously considering an multicomponent model with radiation and non-relativistic matter. 

Most of the models mentioned so far are single-component, and to be considered a worthy alternative, HDE models must include different types of energy densities in order to compare fairly with the mainstream model of cosmology, i.e. the $\Lambda$CDM.
Only a few papers have addressed the study of the interaction of holographic dark energy with both radiation and non-relativistic matter. The vast majority consider only matter, see for example \cite{Zadeh2018,Chakrabarti2018,Zimdahl2007,Zhang2012,Zhang2008}. The present work is an early attempt to investigate a multicomponent holographic dark energy model. 

As an alternative to the hypothesis of the existence of dark energy, it is important to compare HDE models with mainstream cosmology and observational data. The aim of this work is to compare a multicomponent HDE model with the $\Lambda$CDM. Although we recognize the need for observational validation, this is beyond the scope of this work and is the subject of current research.

In this paper, we study how HDE models accommodate a viable cosmology for late times without the consideration of dark energy by using an entropic force term derived from a generic power-law entropy. Furthermore, we consider a correction term necessary to obtain stages of acceleration and deceleration, and derive the temperature of the horizon from the Legendre structure, making our model compatible with thermodynamics. More details on all these considerations can be found in the publications \cite{Zamora2022, Zamora2022b}. We then go one step further by considering a model in which both matter and radiation are present, and we derive the Friedmann, acceleration and continuity equations. We also present a different approach to the deduction of the Friedmann, continuity, and acceleration equations. In previous works \cite{Zamora2022, Komatsu2013b, Zamora2022b} the entropic pressure is introduced in the continuity and acceleration equations, and from there the Friedmann is deduced. Here, we deduce the acceleration equations from the continuity and Friedmann ones. We show that both process takes to the same result. Finally, we present numerical solutions to the equations and discuss the results.

\section{Multicomponent model}

We consider a model under the assumption of a homogeneous, isotropic, and spatially flat universe. Let us consider an entropy that scales with an arbitrary positive power $d\in\mathbb{R^+}$ plus an additional term that depends on a smaller power $0<\Delta<d$.

\begin{equation}
\frac{S}{k_B}=A\left(\frac{r_H}{L_P}\right)^d+E\frac{\left(r_H/L_P\right)^{\Delta}-1}{\Delta},
\label{Snu}
\end{equation}

\nd where $A$ and $E$ are dimensionless constants and $L_P=\sqrt{\hbar G/c^3}$ is the Planck length.
By expressing the subdominant term in this way, we obtain the logarithmic correction presented in \cite{Easson2012} if we take $\Delta\rightarrow0$, since\\
$\lim_{\Delta\rightarrow0}\frac{x^\Delta-1}{\Delta}=\ln{x}$. If we take $E=0$ we also obtain the Bekenstein-Hawking entropy (d=2) and the $\delta=3/2$ entropy \cite{Tsallis2013} (d=3) as special cases.

In order to make our model compatible with thermodynamics, according to Eq. (\ref{key2}), the temperature must scale like $T\propto r_H^{1-d}$. Consequently, we use

\begin{equation}
T=\frac{T_P}{B}\left(\frac{r_H}{L_P}\right)^{1-d},
\end{equation}

\nd as the temperature of the Hubble horizon, where $B$ is a dimensionless factor, and $T_P=\sqrt{\hbar c^5/G k_B^2}$ is the Planck temperature. The entropic force is then given by

\begin{equation}
\begin{split}
F&\equiv-T\frac{dS}{dr_H}=-k_B\frac{d\, A}{B}.\frac{T_P}{L_P}\left[1+\frac{E}{d\,A}\left(\frac{r_H}{L_P}\right)^{\Delta-d}\right]\\
&\equiv-C_d F_P(1+D_{d,\Delta}H^{d-\Delta}),
\end{split}
\end{equation}

\nd where $F_P\equiv k_BT_P/L_P=c^4/G$ is the Planck force, $C_d\equiv d\,A/B$, and $D_{d,\Delta}\equiv E(L_p/c)^{d-\Delta}/(d\,A)$. Therefore, the entropic pressure in the Hubble surface is

\begin{equation}
p_F\equiv\frac{F}{4\pi r_H^2}=-\frac{C_dc^2}{4\pi G}H^2(1+D_{d,\Delta}H^{d-\Delta}).
\end{equation}

Note that, when $d=2$ with $\Delta\rightarrow0$, we obtain the $H^4$ correction term in \cite{Easson2012}, and when $D_{d,\Delta}=0$ we obtain the $H^2$-type models \cite{Easson2011,Zamora2022}.

In our model, we consider a multicomponent universe with non-relativistic matter, for which the energy density has the following dependence:

\begin{equation}
    \rho_m=\frac{3 H_0^2\Omega_m}{8\pi G}\left(\frac{a}{a_0}\right)^{-3},
\end{equation}

\nd and radiation, with density:

\begin{equation}
    \rho_r=\frac{3 H_0^2\Omega_r}{8\pi G}\left(\frac{a}{a_0}\right)^{-4}.
\end{equation}

Remember that the key to the HDE models is that there is no dark energy. Hence the Friedmann equation

\begin{equation}
    \left(\frac{H}{H_0}\right)^2=\frac{\Omega_m}{a^3}+\frac{\Omega_r}{a^4},
    \label{Friedmaneq}
\end{equation}

\nd where the constants $\Omega_m$ and $\Omega_r$ represent the fraction of each type of component and they hold $\Omega_r+\Omega_m=1$, since the assumption of the existence of dark energy is not necessary. Assuming an ideal gas, the total pressure is

\begin{equation}   
    p=\sum_{\omega}p_{\omega}=c^2\sum_{\omega}\omega \rho_{\omega},
\end{equation}

\nd with $\omega$ being the parameter of the equation of state. $\omega$ takes the values $0$ and $1/3$ for non-relativistic matter and radiation, respectively. 

To obtain the continuity equation modified by the entropic force, we replace the effective pressure $p'=p_m+p_r+p_F$ in the usual continuity equation

\begin{equation}
\dot{\rho}+3\frac{\dot{a}}{a}\left(\rho_m+\rho_r+\frac{p'}{c^2}\right)=0,
\label{cont}
\end{equation}

\nd thus arriving to

\begin{equation}        \dot{\rho_r}+\dot{\rho_m}+3\frac{\dot{a}}{a}\left(\rho_r+\rho_m+\frac{p_r}{c^2}+\frac{p_m}{c^2}\right)=\frac{2C_d}{4 \pi G}H^3\left(1+D_{d,\Delta}H^{d-\Delta}\right).
\label{continuityeq}
\end{equation}

From Eq. (\ref{Friedmaneq}) and (\ref{continuityeq}), we can deduce the acceleration equation. To do so, we multiply Eq. (\ref{Friedmaneq}) by $a^2$ and take the time derivative, leading to:

\begin{equation}
    2\ddot{a}\dot{a} = \frac{8\pi G}{3}(\dot\rho_m a^2 + 2 \rho_m a \dot a + \dot\rho_r a^2 + 2\rho_r a \dot a).
\end{equation}

Dividing the last equation by $2\dot a a$, and replacing Eq. (\ref{continuityeq}) we arrive to the modified acceleration equation:
 
\begin{equation}
        \frac{\ddot{a}}{a}=-\frac{4\pi G}{3}\left(\rho_m+\rho_r+\frac{3 p_m}{c^2}+\frac{3p_r}{c^2}\right)+C_d H^2(1+D_{d,\Delta}H^{d-\Delta})
\label{accelerationeq}
\end{equation}

Notice that in the case of a single component universe, following this deduction, we would obtain the equations presented in \cite{Zamora2022b}, obtained in that moment by a different approach. This shows consistency in the model. From the three main eqs. (\ref{accelerationeq}), (\ref{continuityeq}), and (\ref{Friedmaneq}), we obtain

\begin{equation}
    \dot{H}= C_d D_{d,\Delta} H^{d-\Delta+2} + \frac{H^2}{2}\left[2C_d-3(\omega_m+1)\right] - \frac{3}{2}\frac{H_0^2\Omega_r}{a^4}(\omega_r-\omega_m),
\end{equation}

\nd and whit a simple change of variables, we can obtain the differential equation

\begin{equation}
    \frac{dH}{da}= C_d D_{d,\Delta} \frac{H^{d-\Delta+1}}{a} +\frac{H}{2a}\left[2C_d-3(\omega_m+1)\right] - \frac{3}{2}\frac{H_0^2\Omega_r}{Ha^5}(\omega_r-\omega_m).
\label{difeq}
\end{equation}

Eq. (\ref{difeq}) has a known form. First, we rewrite the equation by multiplying $2H$ and changing the variable $H^2=Z$:

\begin{equation}
       \frac{dZ}{da}= \frac{2 C_d D_{d,\Delta}}{a} Z^\frac{d-\Delta+2}{2} +\frac{\left[2C_d-3(\omega_m+1)\right]}{a} Z - 3 \frac{H_0^2\Omega_r}{a^5}(\omega_r-\omega_m).
\end{equation}

This is a special case of the so-called Chini equation

\begin{equation}
    \frac{dy}{dt}= f(t) y(t)^n - g(t)y(t)+h(t),
\end{equation}

\nd which generalizes the Riccati and the Abel equations and is, in general, not solvable but some of its special cases are, see e.g. the book \cite{Kamke1951}. It is known that there is a straightforward recipe for solving the equation in the case that the Chini-invariant is independent of t.

\begin{equation}
    C= n^{-n} f(t)^{-1-n} h(t)^{1-2n}\left( f(t)h'(t) - f'(t)h(t)- nf(t)g(t)h(t)\right)^n.
\end{equation}

In the case examined here, however, this invariant depends on t, so that the method in question is not applicable. We are therefore dealing with an unsolvable equation. There are several ways of dealing with this equation, e.g. working with approximations or solving the equation for different stages (matter-dominated, radiation-dominated and entropic stages). In this paper, we limit ourselves to the analysis of numerical methods to study the viability of the model. More in-depth analyzes are the subject of current research and are currently under development.

\section{Results}

The first thing we want to check is whether the model can accommodate stages of acceleration and deceleration. And whether, in this case, they are similar to those predicted by the $\Lambda$CDM model. To this end, we will study the dependence of the scale factor $a$, with time. We obtain the corresponding differential equation by changing the variables in equation (\ref{difeq}):

\begin{equation}
    \left(\frac{da}{dt}a\right)^2\left[1+\frac{1}{2}[2C_d-3(1+\omega_m)]\right]=\frac{3}{2}\Omega_r(\omega_r-\omega_m)-\frac{C_dD_{d,\Delta}H_0^{d-\Delta}}{a^{d-\Delta-2}}\left(\frac{da}{dt}\right)^{d-\Delta+2}.
    \label{avst}
\end{equation}

\begin{figure}[h!] 
\centering
\begin{subfigure}[b]{0.49\linewidth}
    \centering
    \includegraphics[width=.9\linewidth]{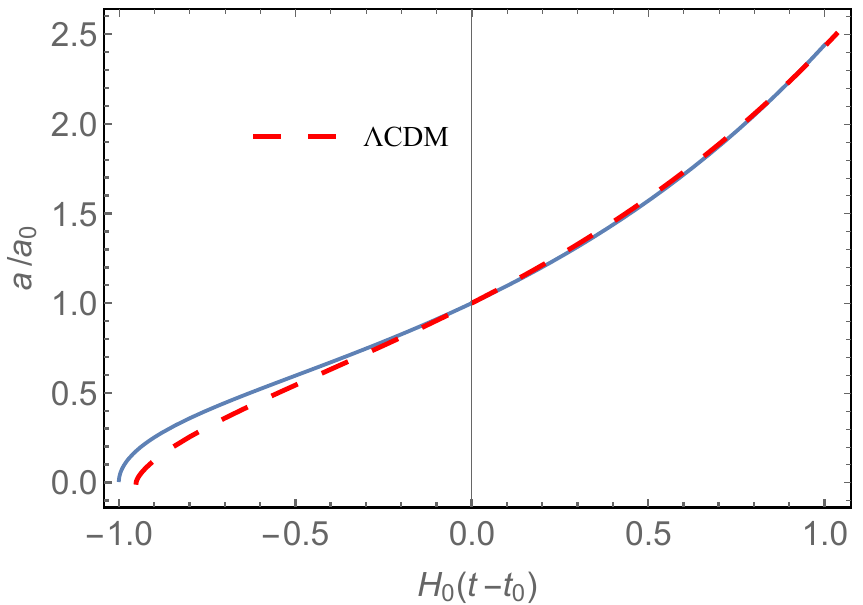}
    \caption{}
\end{subfigure} 
\hfill
\begin{subfigure}[b]{0.49\linewidth}
    \centering
    \includegraphics[width=.9\linewidth]{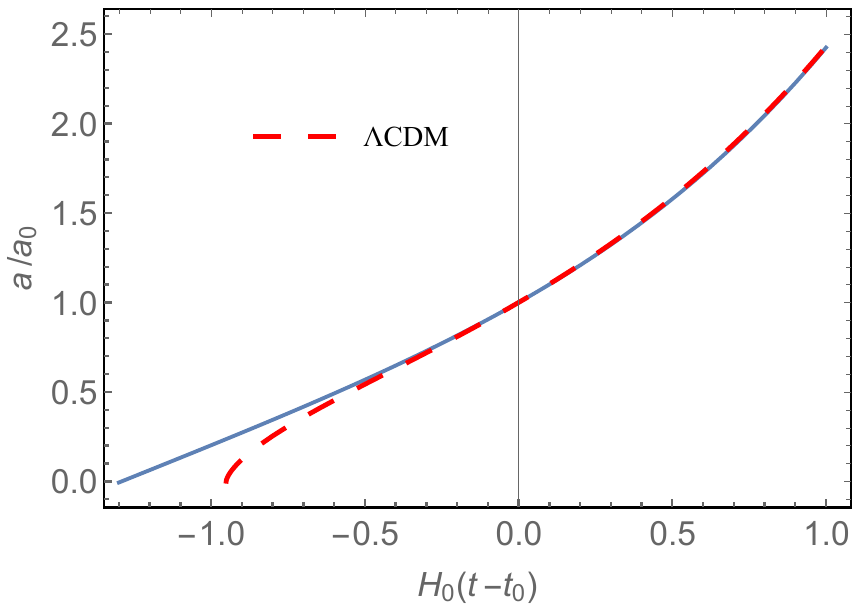}
    \caption{}
\end{subfigure} 
\caption{Scale factor, $a$, as a function of time. Comparison of the numerical solution of the eq. (\ref{avst}) (in blue) with the solution of the $\Lambda$CDM model (red dashed). (a) $d-\Delta=-2$, $C_d=7$ and $D_{d,\Delta}=-0.7$. (b) $d-\Delta=2$, $C_d=-1.5$ and $D_{d,\Delta}=2$   }
\label{fig:1}
\end{figure}

We solve this equation numerically with the function NDsolve from Wolfram Mathematica for different values of the parameters $C_d$, $D_{d,\Delta}$, and $d-\Delta$. We set the other constants to $H_0=67.4\frac{\frac{km}{s}}{mpc}$ (based on the Planck 2018 results \cite{PlanckCollaboration2018}), $\omega_m=0$, and $\omega_r=\frac{1}{3}$. Note that the relationship to the underlying expansion parameters measured (e.g. via the angular diameter distance) depends on the assumed cosmology \cite{PlanckCollaboration2018}. As stated in page 17 of \cite{PlanckCollaboration2018}, after presenting the value of $\Omega_m$: "It is important to emphasize that the values given in Eq. (13) assume the base-$\Lambda$CDM cosmology with minimal neutrino mass. These estimates are highly modeldependent...". Therefore, the obtained values $\Omega_m = 0.3...$ and $\Omega_r = 0.000084$ might not be valid for other models like the present one. In this work, due to the non-existence of dark energy, the relationship $\Omega_r+\Omega_m=1$ must apply. We have decided to set $\Omega_r=2.8\times10^{-4}$, to maintain the same ratio $\Omega_r/\Omega_m$ as in the $\Lambda$CDM model. Of course, other considerations can be made to obtain values for $\Omega_r$ and $\Omega_m$, for example by calculating them from the observational data, but this is beyond the scope of this work. Since $d-\Delta$ is a difference of indices, we expect to find whole numbers. Therefore, we set the parameter $d-\Delta= \pm 1, \pm 2$. We find the numerical solution for different values of $C_d$ and $D_{d,\Delta}$. We found 12 numerical solutions for different combinations of values of the parameters in total. With all these considerations in mind, some results of our model (in blue) together with the $\Lambda$CDM solution are plotted in Fig. (\ref{fig:1}) for comparison.

Several combinations of values of the parameters often gives similar curves, so we show here only two characteristic behaviours. The first thing to notice is that the numerical solution of the model shows different periods of acceleration and deceleration, as expected since we included the correction term. The similarity of the solution of our model with the $\Lambda$CDM model is remarkable, especially in the late time, i.e. in the phase where the 'dark energy' is dominant. This suggests that HDE models are a serious alternative to dark energy models to explain the accelerated expansion of the universe in late time. The behaviour of our model in early time does not match that of the $\Lambda$CDM model, but we can obtain qualitatively similar behaviours depending on the values of the parameters, see for example Fig. (\ref{fig:1}.a). When the parameter $d-\Delta$ takes on positive values, as in Fig. (\ref{fig:1}.b), the change in concavity is not so evident, leading to a much older universe or to behaviour similar to that described in \cite{Zamora2022b}. Negative values of $d-\Delta$ seem more consistent with the $\Lambda$CDM model. However, this is an odd combination of values as $d$ is the dimension and $\Delta$ is the scaling exponent of the correction term. This means that the model needs a correction term that is 'larger" than the main term in order to work. We will discuss this with more detail in the next section.

In addition to the previous results, here we present an early simple comparison between available observational data and the model. The aim of the present paper is not to make a observational validation of the model, but we think is important to present at least a simplified analysis. To do this, we make a change to the variables in equation (\ref{difeq}) and arrive at :

\begin{equation}
    \frac{dH}{dz}=\frac{3}{2}\frac{H_0^2}{H}(1+z)^3\Omega_r(\omega_r-\omega_m)-\frac{C_dD_{d,\Delta}}{1+z}H^{d-\Delta+1}-\frac{[2C_d-3(1+\omega_m)]}{2}\frac{H}{1+z}.
\end{equation}

\begin{figure}[h!] 
\centering
\begin{subfigure}[b]{0.49\linewidth}
    \centering
    \includegraphics[width=0.9\linewidth]{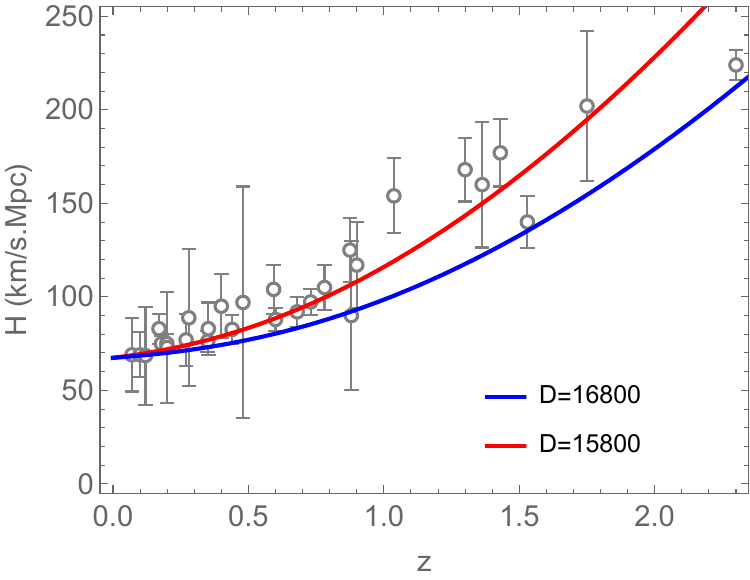}
    \caption{}
    \label{fig:enter-label1}
\end{subfigure} 
\hfill
\begin{subfigure}[b]{0.49\linewidth}
    \centering
    \includegraphics[width=0.9\linewidth]{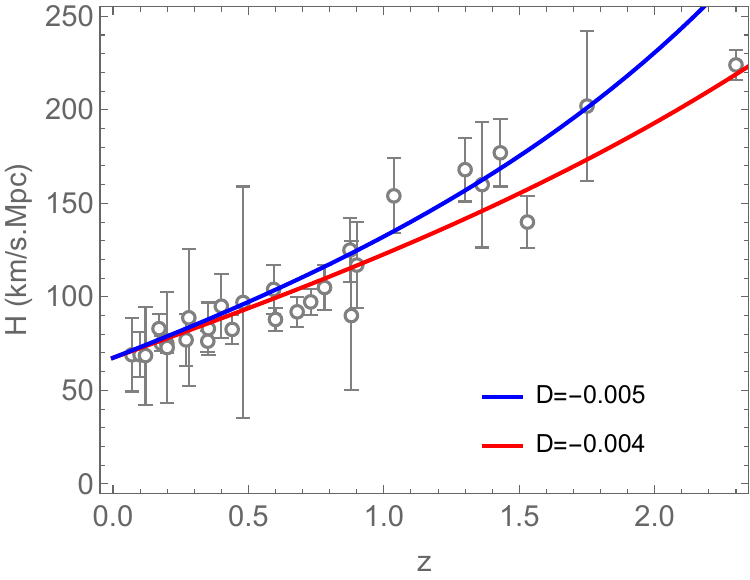}
    \caption{}
    \label{fig:enter-label2}
\end{subfigure} 
\caption{Hubble parameter $H$ versus redshift $z$. The open circle with bars are data points taken from Table 1 in \cite{Pradhan2021}. In all cases, the value of $H_0$ is set to be $67.4 km/s/Mpc$ based on the Planck 2018 results \cite{PlanckCollaboration2018}. Red and blue curves correspond to different combinations of values of the parameters of the model. (a) $d-\Delta=-2$, $C_d=-0.5$ and $D_{d,\Delta}=15800$ (red line), $D_{d,\Delta}=16800$ (blue line). (b)  $d-\Delta=1$, $C_d=1$, and $D_{d,\Delta}=-0.004$ (red line), $D_{d,\Delta}=-0.005$ (blue line).}
\end{figure}

\begin{figure}[h]
    \centering
    \includegraphics[width=0.5\linewidth]{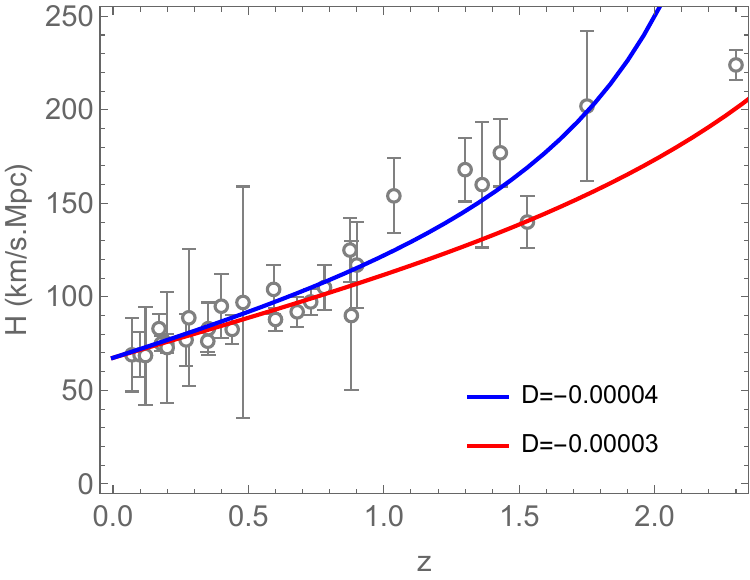}
    \caption{Hubble parameter $H$ versus redshift $z$. The open circle with bars are data points taken from Table 1 in \cite{Pradhan2021}. In all cases, the value of $H_0$ is set to be $67.4 km/s/Mpc$ based on the Planck 2018 results \cite{PlanckCollaboration2018}. Red and blue curves correspond to different combinations of values of the parameters of the model: $d-\Delta=2$, $C_d=1$ and $D_{d,\Delta}=-0.00003$ (red line), and $D_{d,\Delta}=-0.00004$ (blue line).}
    \label{fig:enter-label3}
\end{figure}

In figures (\ref{fig:enter-label1}), (\ref{fig:enter-label2}) and (\ref{fig:enter-label3}) we have plotted the Hubble parameter $H$ as a function of redshift $z$ using the data points from Table 1 in \cite{Pradhan2021}, which in turn come from Supernova SN1a databases.
In Fig. (\ref{fig:enter-label1}) we assume $d-\Delta=-2$ and determine the constants $C_d$ and $D_{d,\Delta}$ by performing a sweep of the values. We show the plots of the combinations of the values $d-\Delta=-2$, $C_d=-0.5$ and $D_{d,\Delta}=15800$ (red line) and $D_{d,\Delta}=16800$ (blue line). In Fig. (\ref{fig:enter-label2}) we show solutions with the combination $d-\Delta=1$, $C_d=1$, and $D_{d,\Delta}=-0.004$ (red line), $D_{d,\Delta}=-0.005$ (blue line).

Finally, we show the combination $d-\Delta=2$, $C_d=1$ and $D_{d,\Delta}=-0.00003$ (red), $D_{d,\Delta}=-0.00004$ (blue) in Fig. (\ref{fig:enter-label3}). We see that the model has a satisfactory fit to the data points in all cases.

\section{Discussion and conclusions}
In this paper, we have explored a multicomponent holographic dark energy model, also known as entropic cosmology, as an alternative explanation for the late accelerated expansion of the universe. The key idea behind this model is to attribute the acceleration to entropy associated with information storage in the universe, eliminating the need for an ad-hoc dark energy component. The model includes a subdominant power-law term, introducing a correction to the main entropy term. A noteworthy observation is that for the model to predict similar behaviour to $\Lambda$ CDM model, the parameter $d-\Delta$ has to be negative. This implies that the correction term must be larger than the main term, emphasizing the importance of the correction in driving the dynamics. The necessity of negative values for $d-\Delta$ requires careful consideration of the physical implications. Maybe the cosmological entropy of the Hubble horizon is not a known entropy yet, such as Bekenstein-Hawking, Boltzmann-Gibbs, or Tsallis entropy, but a new different one, with a similar form to the correction term here presented.
A second problem is that several parameters in the model lack clear physical interpretation, raising concerns about the model's theoretical foundation. All the new parameters in this model come from the entropy considered. The parameter $d-\Delta$ has a clear interpretation, but $C_d$ and $D_{d,\Delta}$ no. They are a combination of the exponents $d$, $\Delta$, and of the coefficients $A$, $E$, and $B$ escorting the entropy and temperature. In the simplest case in which we take $A = B = E = 1$, $C_d$ and $D_{d,\Delta}$ depend only in physical constant, acquiring in this way clear physical interpretation. However, in this case, the model loses all its predictive power, since the parameters are necessary to fit the available observational data. The model's predictions need to undergo observational verification, acknowledging that this is an essential aspect yet to be explored in detail. Comparison with available observational data, particularly Hubble parameter and luminosity distance from supernovae, is necessary to establish the model's credibility. Observational validation is crucial, and the model's compatibility with a broader range of observational data remains to be investigated.

The holographic dark energy model aims to address the shortcomings of the $\Lambda$CDM model, which relies on a cosmological constant and introduces the mysterious dark energy. The model can potentially explain different stages of acceleration and deceleration without invoking dark energy and produces results comparable to the $\Lambda$CDM model in late-time expansion. This is an advantage since the model provides a unique and concrete physical interpretation of the late accelerated expansion through entropy and the holographic principle. The correction term allows for flexibility in describing different periods of acceleration and deceleration.

In conclusion, while the holographic dark energy model presents intriguing possibilities and potential advantages over the traditional 
$\Lambda$CDM model, careful scrutiny, theoretical refinement, and observational testing are essential for establishing its validity and addressing the identified concerns. Further research and detailed observational comparisons will be crucial in determining the true merits and limitations of this cosmological approach.
%Same structure as abstract but expanded

\section*{Acknowledgment}
This work received financial support from CONICET, and CIN (Argentinian agencies).

\end{document}